\documentclass[iop]{emulateapj}

\newcommand{\msolar}{M$_{\odot}$}

\newcommand{\ml}{M$_{\odot}$ yr$^{-1}$}

\usepackage{subfigure}
\usepackage{multirow}
\usepackage[dvips]{color}
\usepackage{epstopdf}
\definecolor{Mygrey}{gray}{0.6}

\begin{document}

\title{The Candidate Progenitor of the Type II\MakeLowercase{n} SN 2010\MakeLowercase{jl} is Not an Optically Luminous Star}
\shorttitle{Progenitor of SN 2010jl}
\author{Ori D. Fox\altaffilmark{1,2}, Schuyler D. Van Dyk\altaffilmark{3}, Eli Dwek\altaffilmark{4}, Nathan Smith\altaffilmark{5}, Alexei V. Filippenko\altaffilmark{6}, Jennifer Andrews\altaffilmark{5}, Richard G. Arendt\altaffilmark{7}, Ryan~J.~Foley\altaffilmark{8,9,10}, Patrick L. Kelly\altaffilmark{6}, Adam A. Miller\altaffilmark{11,12,13}, and Isaac Shivvers\altaffilmark{6}}
\altaffiltext{1}{Space Telescope Science Institute, 3700 San Martin Drive, Baltimore, MD 21218, USA.}
\altaffiltext{2}{ofox@stsci.edu.}
\altaffiltext{3}{IPAC/Caltech, Mailcode 100-22, Pasadena, CA 91125, USA.}
\altaffiltext{4}{Astrophysics Science Division, NASA Goddard Space Flight Center, Mail Code 665, Greenbelt, MD 20771, USA.}
\altaffiltext{5}{Steward Observatory, 933 N. Cherry Ave., Tucson, AZ 85721, USA.}
\altaffiltext{6}{Department of Astronomy, University of California, Berkeley, CA 94720-3411, USA.}
\altaffiltext{7}{CRESST/UMBC/GSFC Code 665, NASA/GSFC, Greenbelt MD, 20771, USA.}
\altaffiltext{8}{Department of Astronomy and Astrophysics, University of California, Santa Cruz, CA 95064, USA.}
\altaffiltext{9}{Astronomy Department, University of Illinois at Urbana-Champaign, 1002 W.\ Green Street, Urbana, IL 61801, USA.}
\altaffiltext{10}{Department of Physics, University of Illinois at Urbana-Champaign, 1110 W.\ Green Street, Urbana, IL 61801, USA.}
\altaffiltext{11}{Jet Propulsion Laboratory, 4800 Oak Grove Drive, MS 169-506, Pasadena, CA 91109, USA.}
\altaffiltext{12}{California Institute of Technology, Pasadena, CA 91125, USA.}
\altaffiltext{13}{Hubble Fellow.}
\begin{abstract}

The nature of the progenitor star (or system) for the Type IIn supernova (SN) subclass remains uncertain.  While there are direct imaging constraints on the progenitors of at least four Type IIn supernovae, one of them being SN 2010jl, ambiguities remain in the interpretation of the unstable progenitors and the explosive events themselves.  A blue source in pre-explosion {\it HST}/WFPC2 images falls within the 5$\sigma$~astrometric error circle derived from post-explosion ground-based imaging of SN 2010jl.  At the time the ground-based astrometry was published, however, the SN had not faded sufficiently for post-explosion {\it HST} follow-up observations to determine a more precise astrometric solution and/or confirm if the pre-explosion source had disappeared, both of which are necessary to ultimately disentangle the possible progenitor scenarios.  Here we present {\it HST}/WFC3 imaging of the SN 2010jl field obtained in 2014 and 2015, when the SN had faded sufficiently to allow for new constraints on the progenitor.  The SN, which is still detected in the new images, is offset by $0{\farcs}099 \pm 0{\farcs}008$~(24 $\pm$~2 pc) from the underlying and extended source of emission that contributes at least partially, if not entirely, to the blue source previously suggested as the candidate progenitor in the WFPC2 data.  This point alone rules out the possibility that the blue source in the pre-explosion images is the exploding star, but may instead suggest an association with a young ($<5-6$ Myr) cluster and still argues for a massive ($>30$ \msolar) progenitor.  We obtain new upper limits on the flux from a single star at the SN position in the pre-explosion WFPC2 and {\it Spitzer}/IRAC images that may ultimately be used to constrain the progenitor properties.
\end{abstract}


\section{Introduction}
\label{sec:intro}
Type IIn supernovae (SNe~IIn; see \citealt{filippenko97} for a review) are core-collapse explosions whose spectra are characterized by relatively narrow lines \citep{schlegel90} which are not associated with the supernova (SN) explosion itself, but rather with a dense circumstellar shell (CS) produced by pre-SN mass loss.  The nature of the progenitor star (or system) remains uncertain and need not be limited to a single solution.  SNe IIn exhibit a range of light-curve characteristics \citep[e.g.,][and references within]{taddia13} and derived pre-SN mass-loss rates ($10^{-4} - 10^{-1}$~\ml; e.g., \citealt{smith09ip, fox09,fox11,moriya13}).  Galactic analogs with such mass-loss rates include anything from extreme self-obscured red supergiants (RSGs) to luminous blue variables (LBVs), each of which present further questions of their own (see \citealt{smith14all} for a review).  

Direct imaging constraints on Type IIn progenitors are limited to only a handful of cases: SNe 1961V, 2005gl, 2009ip, 2010jl, and 2015bh  \citep{goodrich89,filippenko95,vandyk02,smith1161v,kochanek11,vandyk12, galyam07,galyam09,foley11,smith11jl,elias-rosa16,thoene16,smith16}.  While each case has a unique set of caveats that should be carefully considered, a blue and/or overly luminous source at the position of each of these SNe was discovered in the pre-explosion images that can be considered consistent with a luminous, high-mass star, which is most typically labeled a LBV [also see the case of the pre-explosion outbursts in the Type Ibn SN 2006jc  \citep{foley07,pastorello07} and Type IIn-P 2011ht \citep{fraser13}].

\begin{figure*}[t]
\begin{center}
\epsscale{0.33}
\subfigure{\label{f1a} \plotone{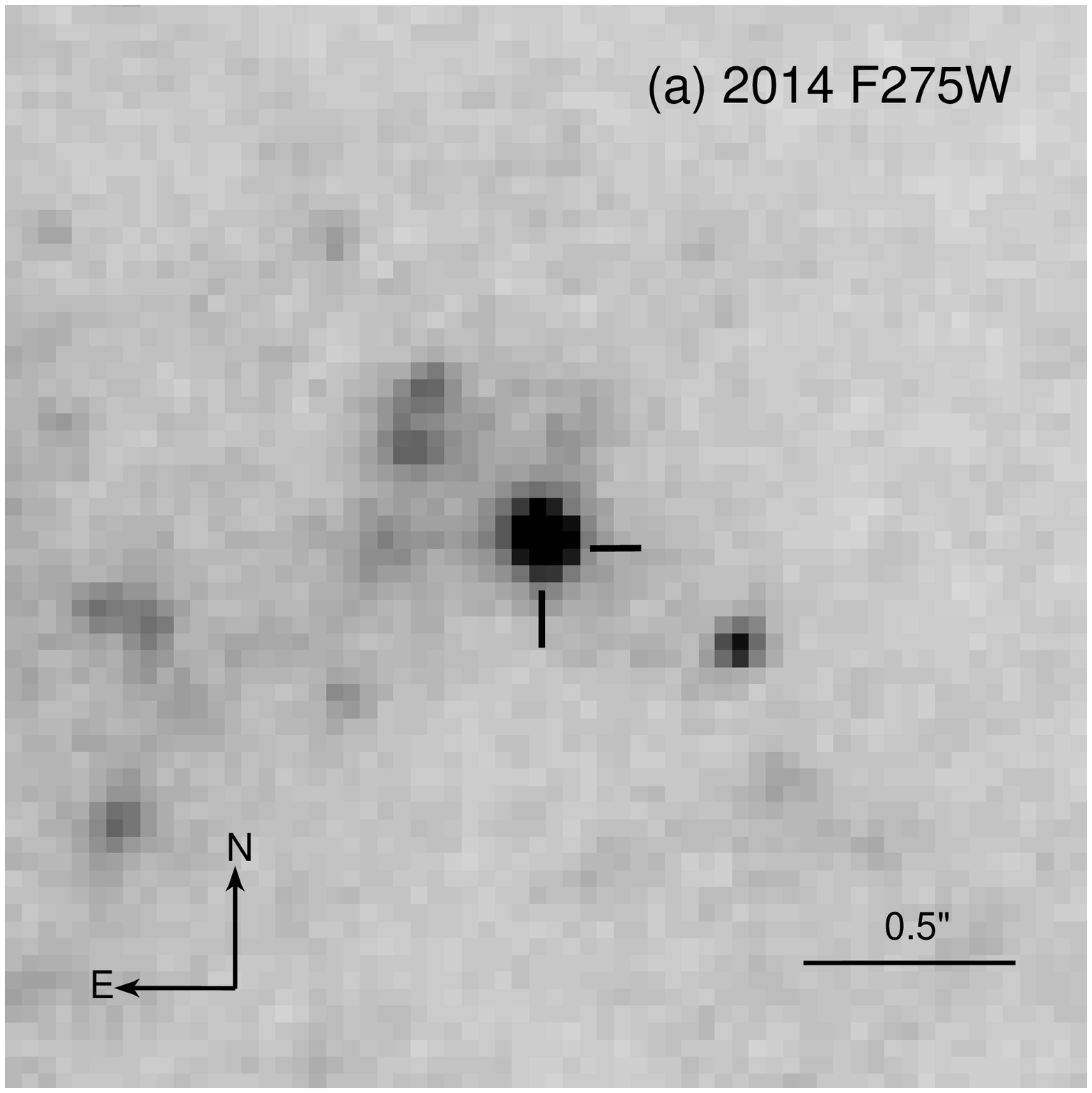}}
\subfigure{\label{f1c} \plotone{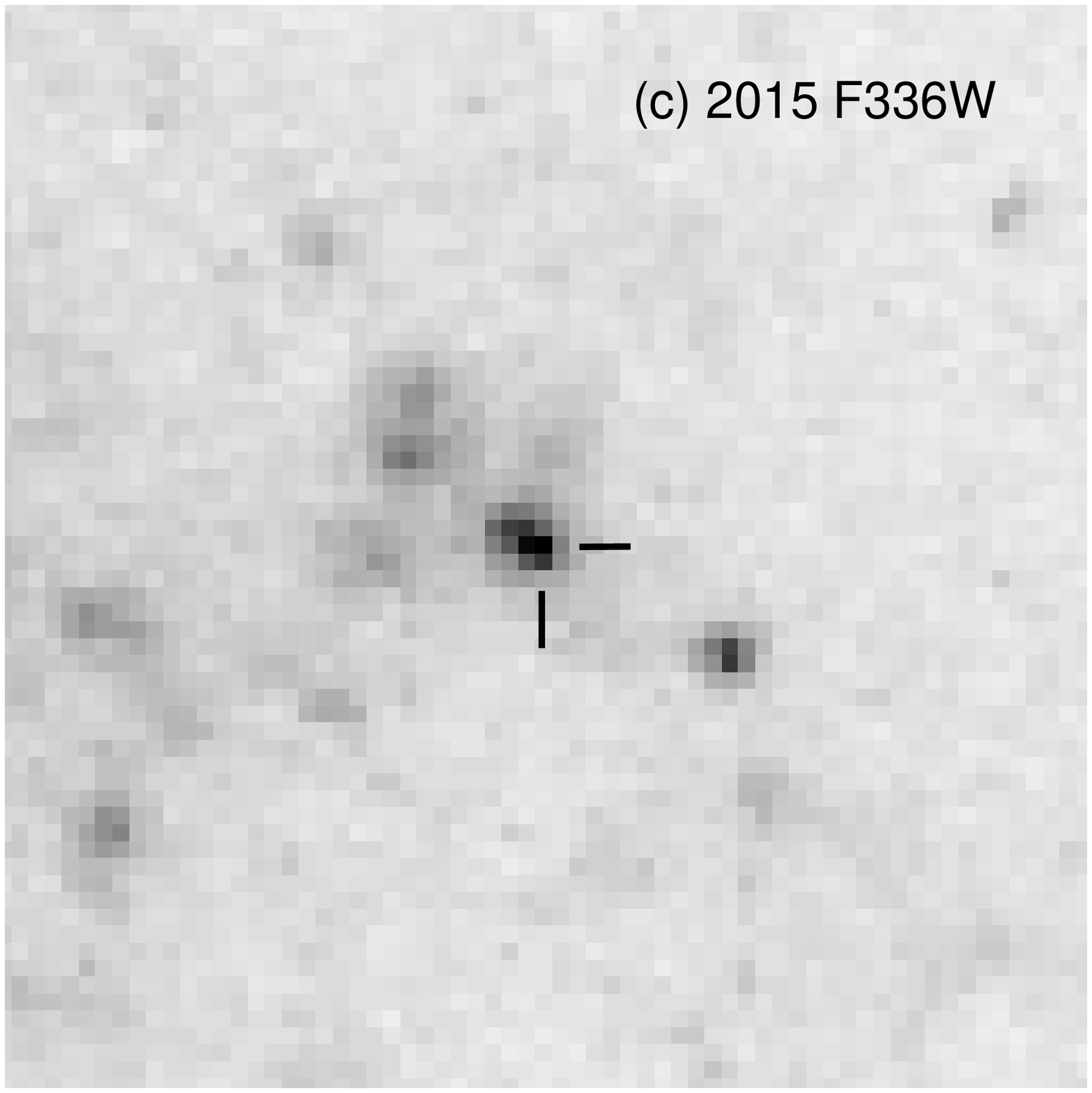}}
\subfigure{\label{f1e} \plotone{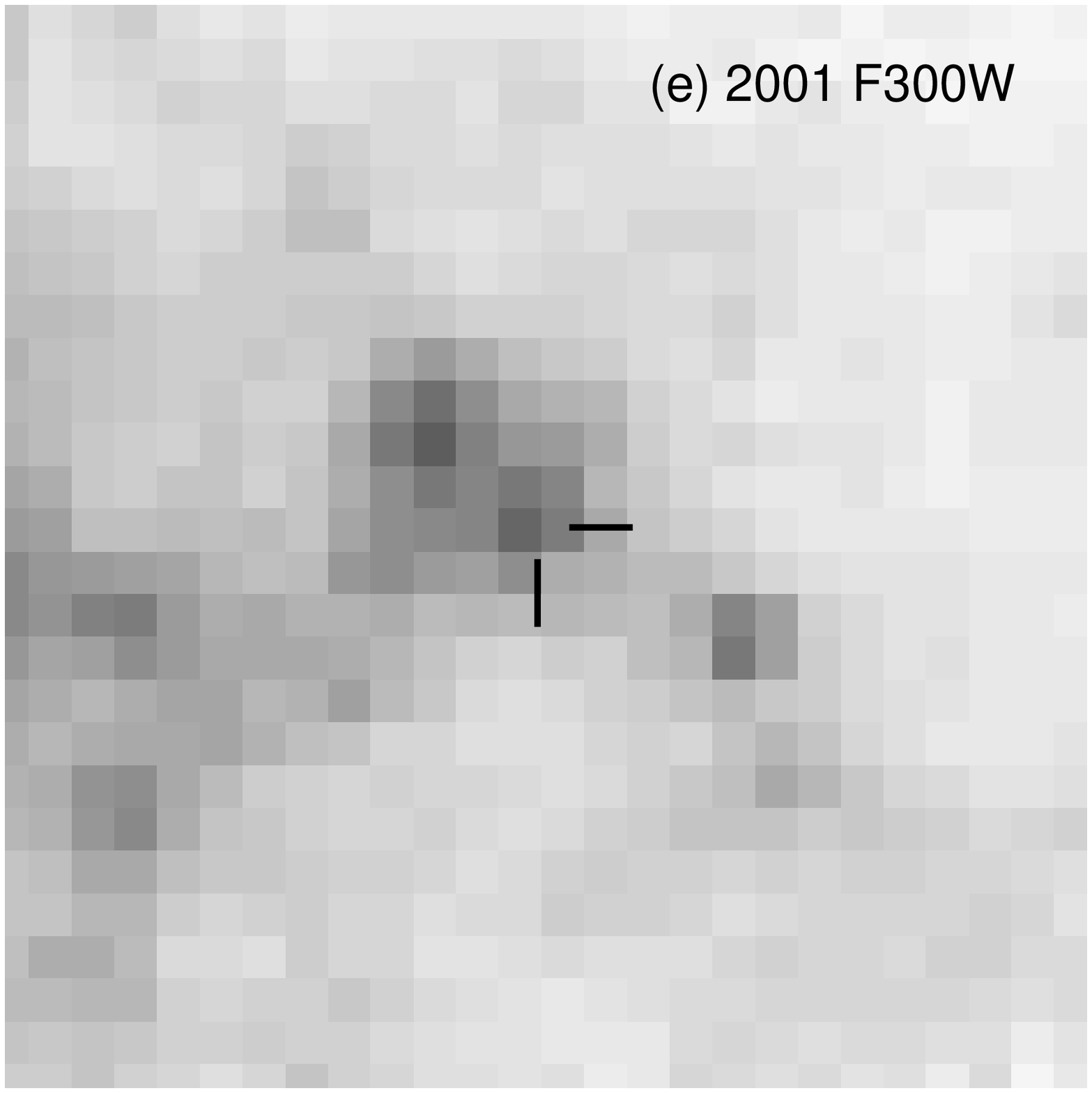}}\\
\subfigure{\label{f1b} \plotone{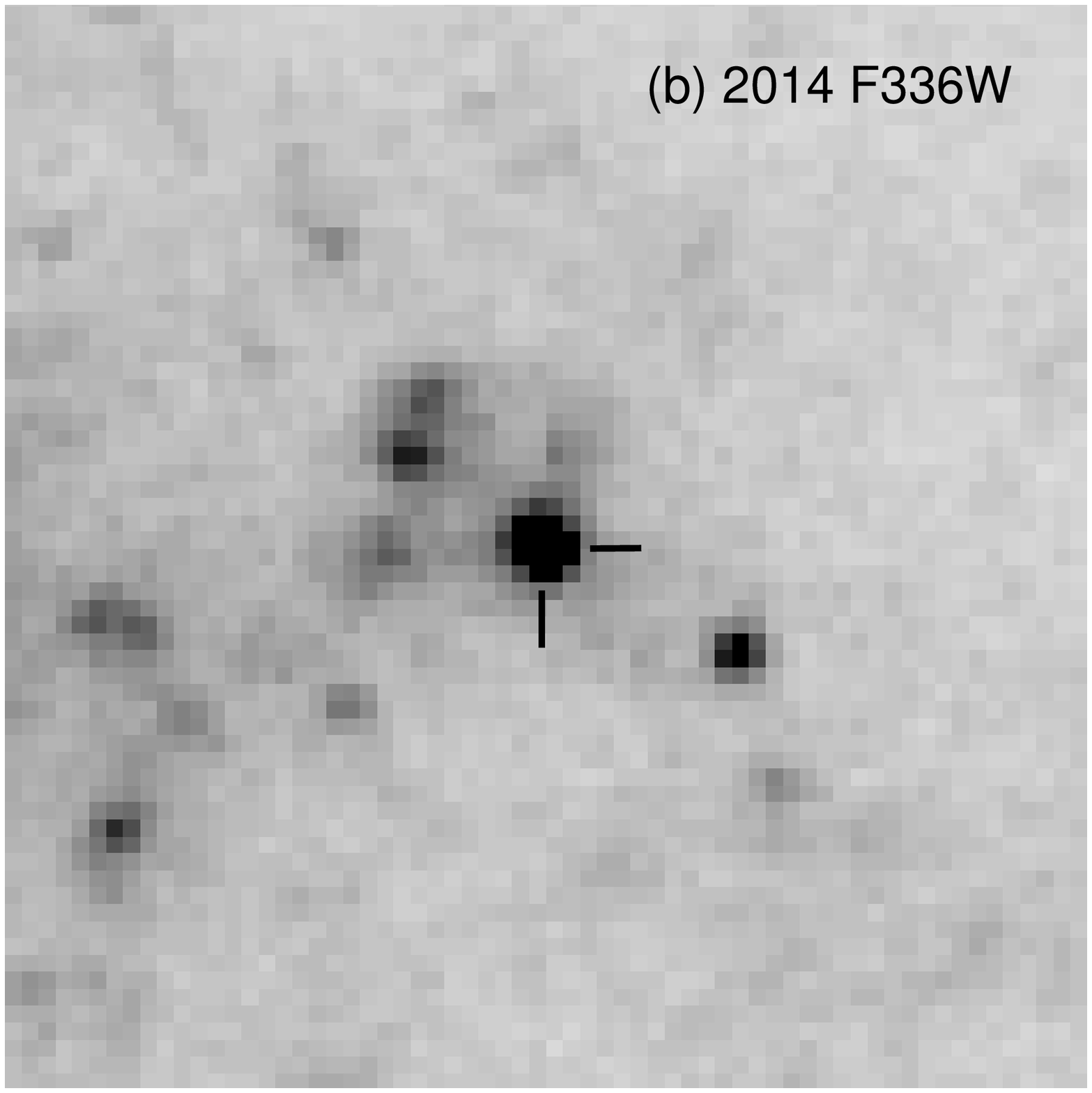}}
\subfigure{\label{f1d} \plotone{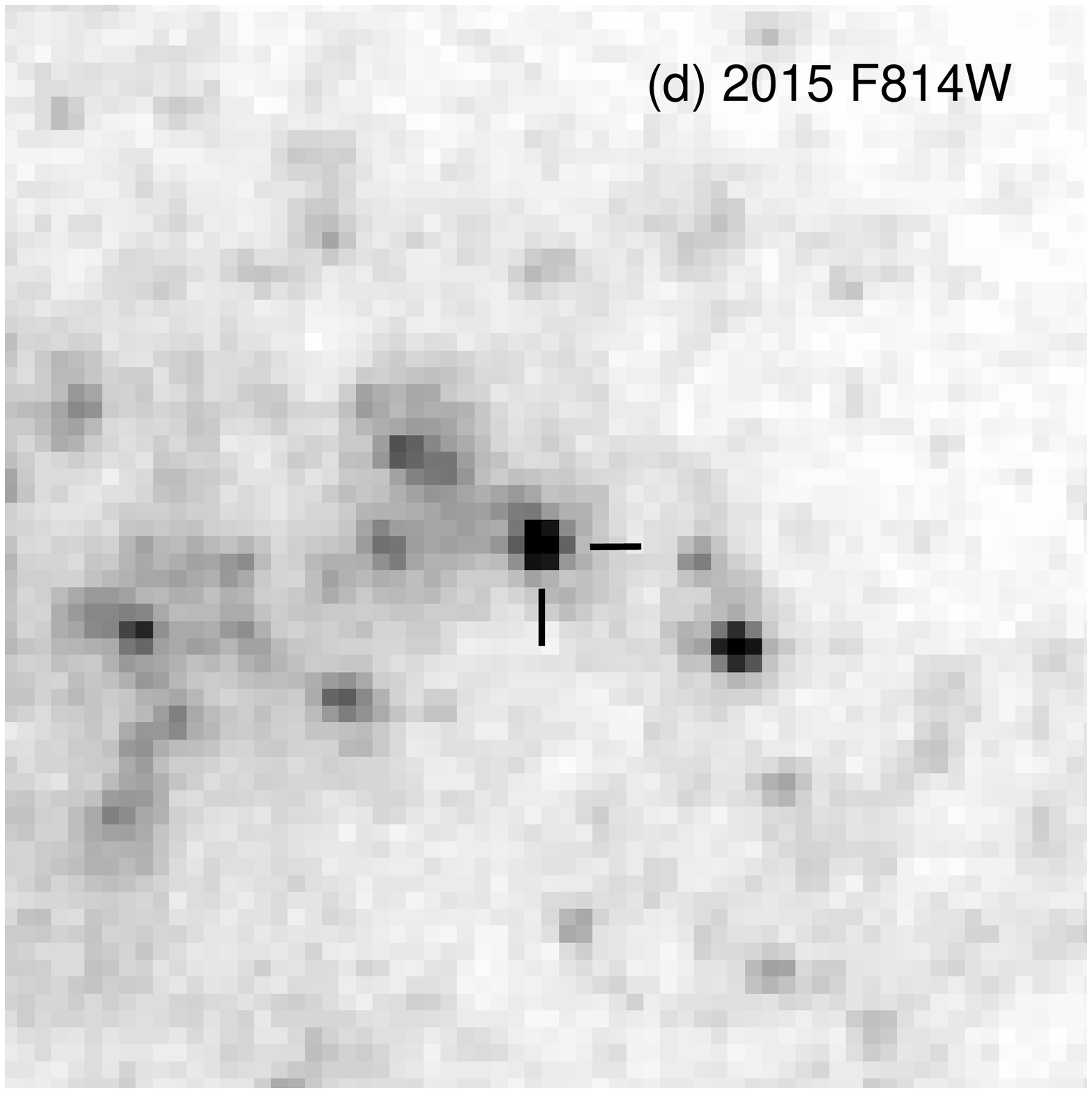}}
\subfigure{\label{f1f} \plotone{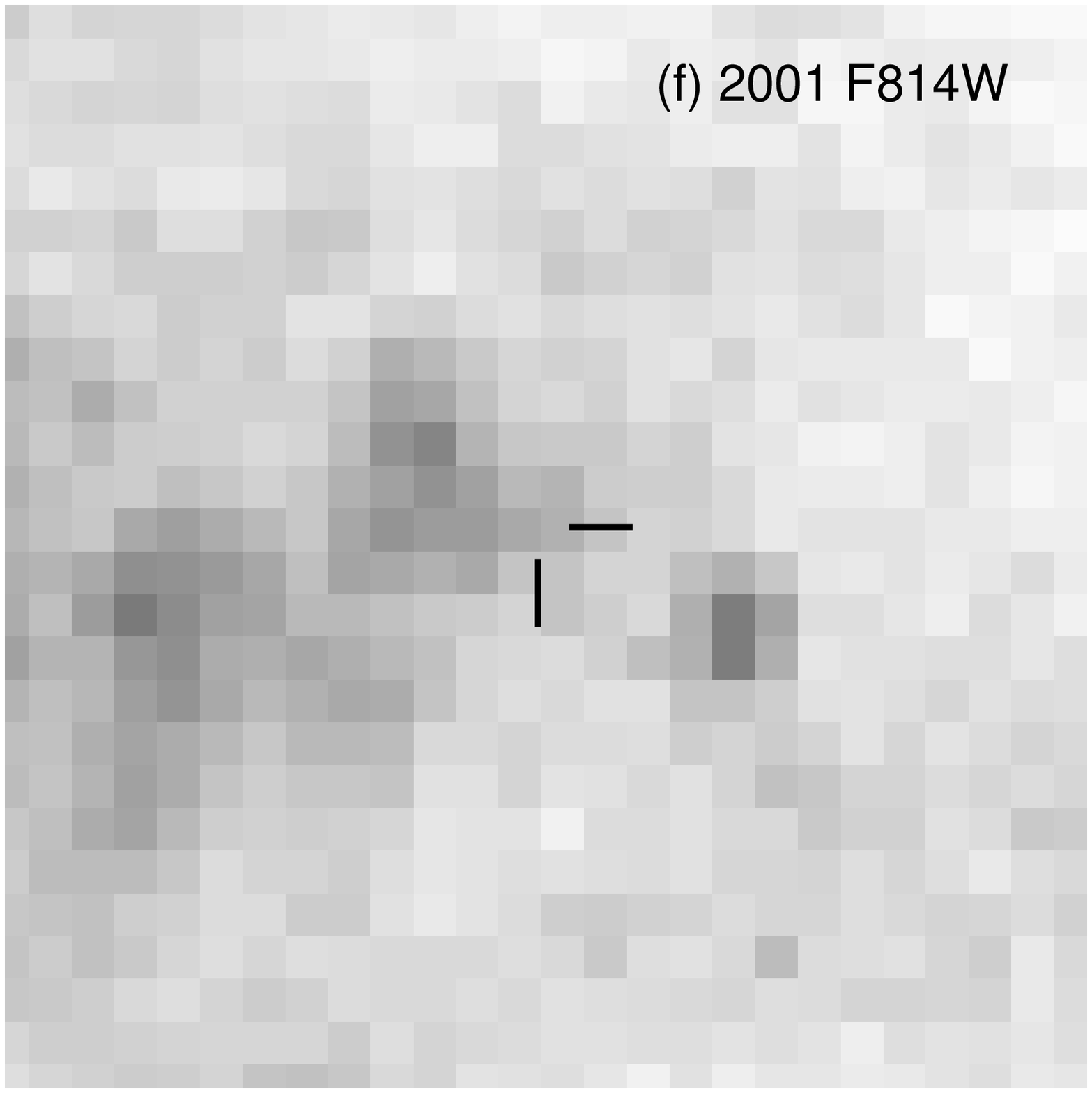}}
\caption{{\it HST} WFC3 post-explosion images of SN 2010jl (a)-(d) and WFPC2 pre-explosion (e)-(f).  The WFC3/F336W image from 2015 (Figure \ref{fig1}(c)) shows that SN 2010jl is demonstrably offset from an underlying and extended source of emission that contributes at least partially, if not entirely, to the blue source identified as the progenitor in the WFPC2 data.  The new SN coordinates are marked in each of the images.  
}
\label{fig1}
\end{center}
\end{figure*}

At the time of writing this article, however, sufficient ambiguities exist around each scenario to suggest it is still premature to claim a quiescent Type IIn progenitor has been definitively discovered.  For example, SN 2009ip has indeed faded below the brightness of the detected progenitor, but it is uncertain if that progenitor was in its quiescent state \citep{thoene15,smith16}.  SN 2005gl also faded below pre-explosion luminosities, but the limits constrain this dimming to only $\gtrsim1.5$ mag.  The deep limits for SN 1961V show that it dimmed by $\gtrsim5.5$ mag, but a debate still exists concerning whether SN 1961V was a true SN or a nonterminal eruption with a fainter, surviving source \citep[see][and references therein]{smith11impostor,kochanek11,vandyk12}.  The uncertainty in all of these cases can be summarized by the fact that LBVs are known to undergo quiescent and eruptive stages that can differ by $>3$ mag \citep{wolf92}.

These ambiguities aside, the lack of massive-star progenitor detections is puzzling.  To complicate the interpretation even more, \citet{habergham14}~find that SNe~IIn do not trace star formation in galaxies, suggesting they are not likely associated with the most massive stars (e.g., LBVs).  \citet{smith15} and \citet{smith16} go on to show that LBVs are more isolated from O stars than predicted by single-star evolution models.  Instead, these authors propose an alternative progenitor scenario that utilizes mass gainers in Roche-lobe overflow (RLOF; but see \citealt{humphreys16} for a contrasting interpretation of how to subdivide the sample of LBVs.)

The Type IIn SN 2010jl was discovered in host galaxy UGC 5189A on 2010 November 3.52 (UT dates are used throughout this paper) by \citet{newton10}.  \citet{smith11jl} identified a blue source in pre-explosion {\it Hubble Space Telescope}~({\it HST}) Wide-Field Planetary Camera 2 (WFPC2) images that falls within the 5$\sigma$~astrometric error circle ($\sigma \approx 0{\farcs}05$) derived from post-explosion ground-based imaging.  The blue color of this pre-explosion source is consistent with either (1) a massive young ($<5-6$ Myr) star cluster, (2) a luminous blue star with an apparent temperature around 14,000 K, (3) a star caught during a bright outburst similar to those of LBVs, or (4) a combination of the above.  At the time of that publication, no ground-based adaptive optics observations were acquired and the SN had not faded sufficiently for post-explosion {\it HST} follow-up images to determine a more precise astrometric solution or confirm if the pre-explosion source had disappeared.

This paper presents observations of SN 2010jl obtained with {\it HST} Wide-Field Camera 3 (WFC3) $\sim5$ yr post-explosion, at which time the SN has faded enough to obtain sufficiently accurate astrometry (especially considering the WFC3 plate scale offers a resolution that exceeds that of WFPC2 by a factor of $\sim2.5$).  Here we present those observations to determine if the blue source identified in pre-explosion images was the progenitor or part of a massive star cluster.  Section \ref{sec:obs} presents the observations and an analysis of the astrometry.  Section \ref{sec:con} summarizes our conclusions.

\begin{figure*}[t]
\begin{center}
\epsscale{0.55}
\subfigure{\label{f2a} \plotone{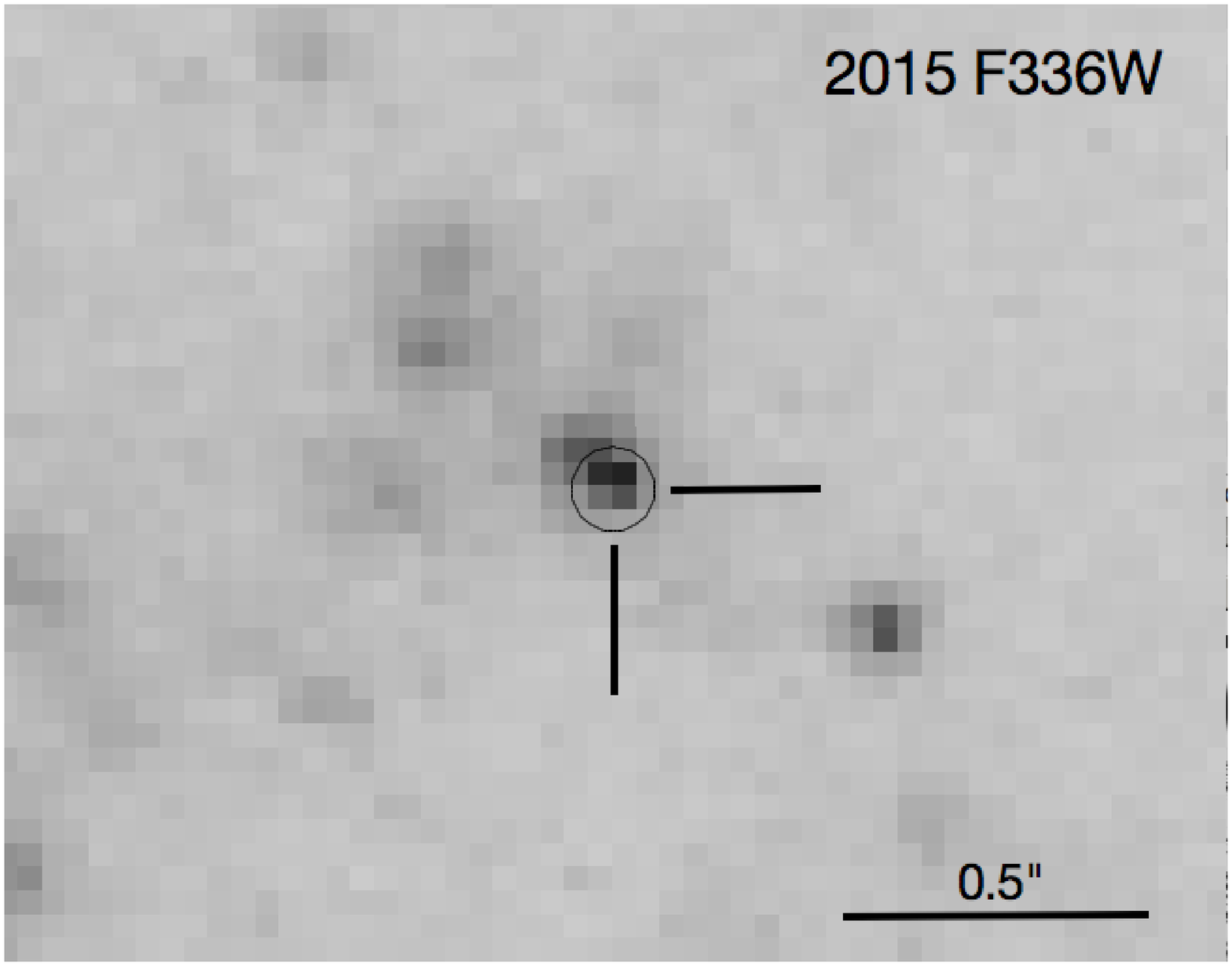}}
\subfigure{\label{f2b} \plotone{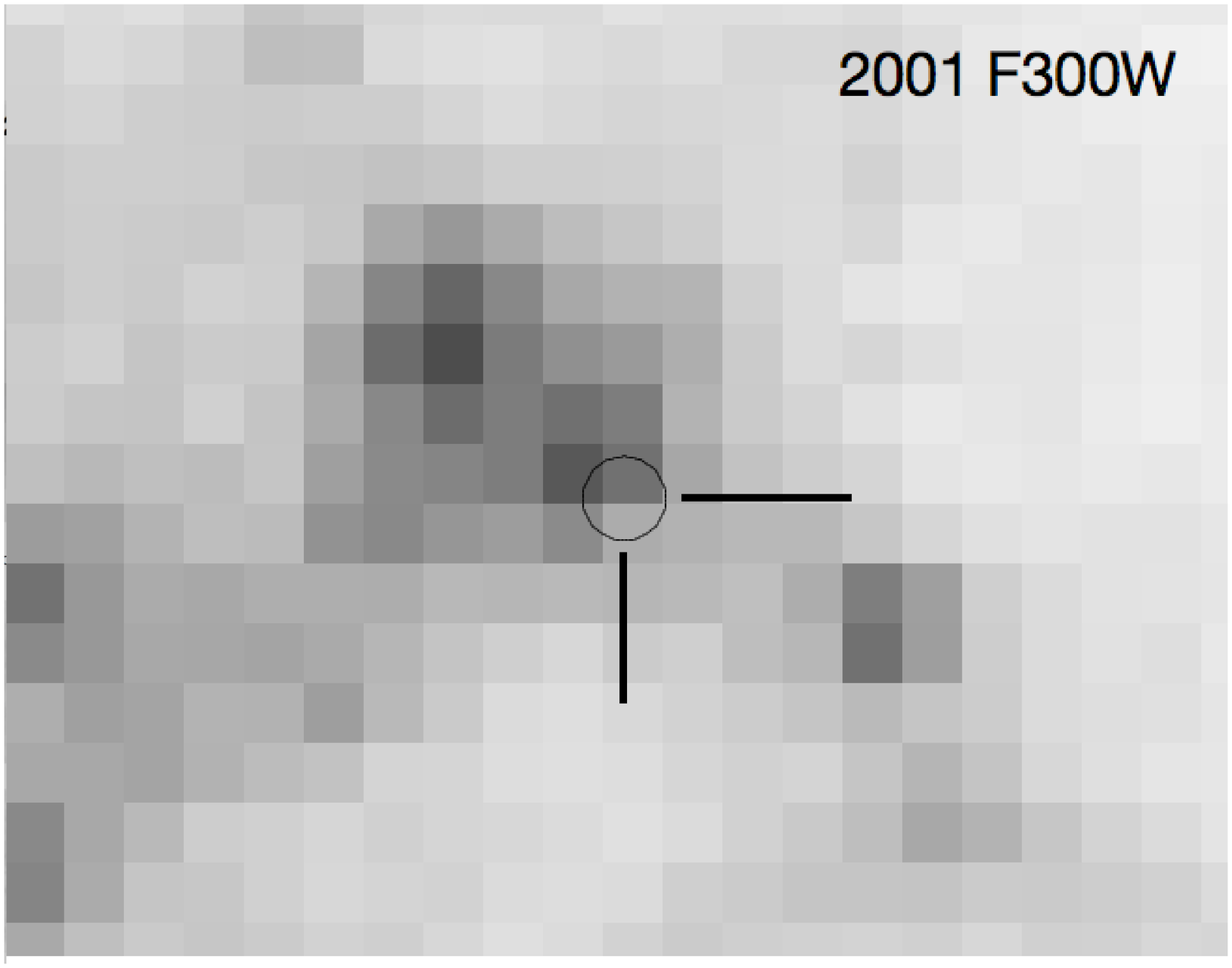}}
\caption{Same as Figure \ref{fig1}(c) and (e), but zoomed in on the new SN position in both the post- and pre-explosion images.
}
\label{fig2}
\end{center}
\end{figure*}

\begin{deluxetable}{ l c c c c}
\tablewidth{0pt}
\tabletypesize{\normalsize}
\tablecaption{$HST$/WFC3 Photometry (mag) of SN 2010jl and its Progenitor \label{tab1}}
\tablecolumns{5}
\tablehead{
\colhead{UT Date} & \colhead{Epoch} & \colhead{F275W} & \colhead{F300W/F336W} &  \colhead{F814W} 
}
\startdata
20091110 & $-358$ &  --- & $<$25.7 (0.1) & $<$25.6 (0.1) \\
20141110 & 1468 & 19.82 (0.01) & 20.86 (0.02) & --- \\
20151010 & 1802 & --- & 21.57 (0.03) & 22.14 (0.02)
\enddata
\end{deluxetable}

\section{Observations}
\label{sec:obs}
\subsection{HST/STIS}
SN 2010jl was observed  with the Space Telescope Imaging Spectrograph (STIS) by program GO-12242 (PI R.~Kirshner) in the 50CCD aperture with the MIRVIS filter/grating on 2011 January 23 for 120~s, but the image of the SN is saturated. We nonetheless attempted to measure a SN position relative to the archival WFPC2 data, but it was very difficult to to establish a positional centroid and therefore the measurement was deemed not useful. It turns out, though, that the position we estimated is ~$0{\farcs}02$ from the actual SN position.

\subsection{HST/WFC3}

SN 2010jl was observed with the {\it HST}/WFC3 UVIS channels as part of programs GO-13341 (PI S. Van Dyk) and GO-14149 (PI A. Filippenko), shown in Figures \ref{fig1}(a)-(d).  These data represent the first unsaturated {\it HST} images of SN 2010jl obtained post-explosion.  The individual WFC3 ``flt'' images were first corrected for charge-transfer efficiency losses using the scripts available online\footnote{http://www.stsci.edu/hst/wfc3/tools/cte\_tools.}.  The resulting ``flc'' images then had cosmic-ray hits masked by running them through AstroDrizzle in PyRAF.  Photometry was extracted from the  individual WFC3 ``flt'' images in all bands using DOLPHOT v2.0 \citep{dolphin00}.  We adopted a number of the DOLPHOT input parameters recommended by \citet{dalcanton09} and \citet{radburn-smith11}, as appropriate to complex backgrounds in nearby galaxies; in particular, we used FitSky=3 and RAper=10 (although we set SkipSky=1), as well as the Anderson point-spread function (PSF) library \citep{anderson06}.  Aperture corrections were applied.  The resulting magnitudes in the WFC3 flight system (Vegamag) are listed in Table~\ref{tab1}.

\begin{figure*}[t]
\begin{center}
\epsscale{0.5}
\subfigure{\label{f3a} \plotone{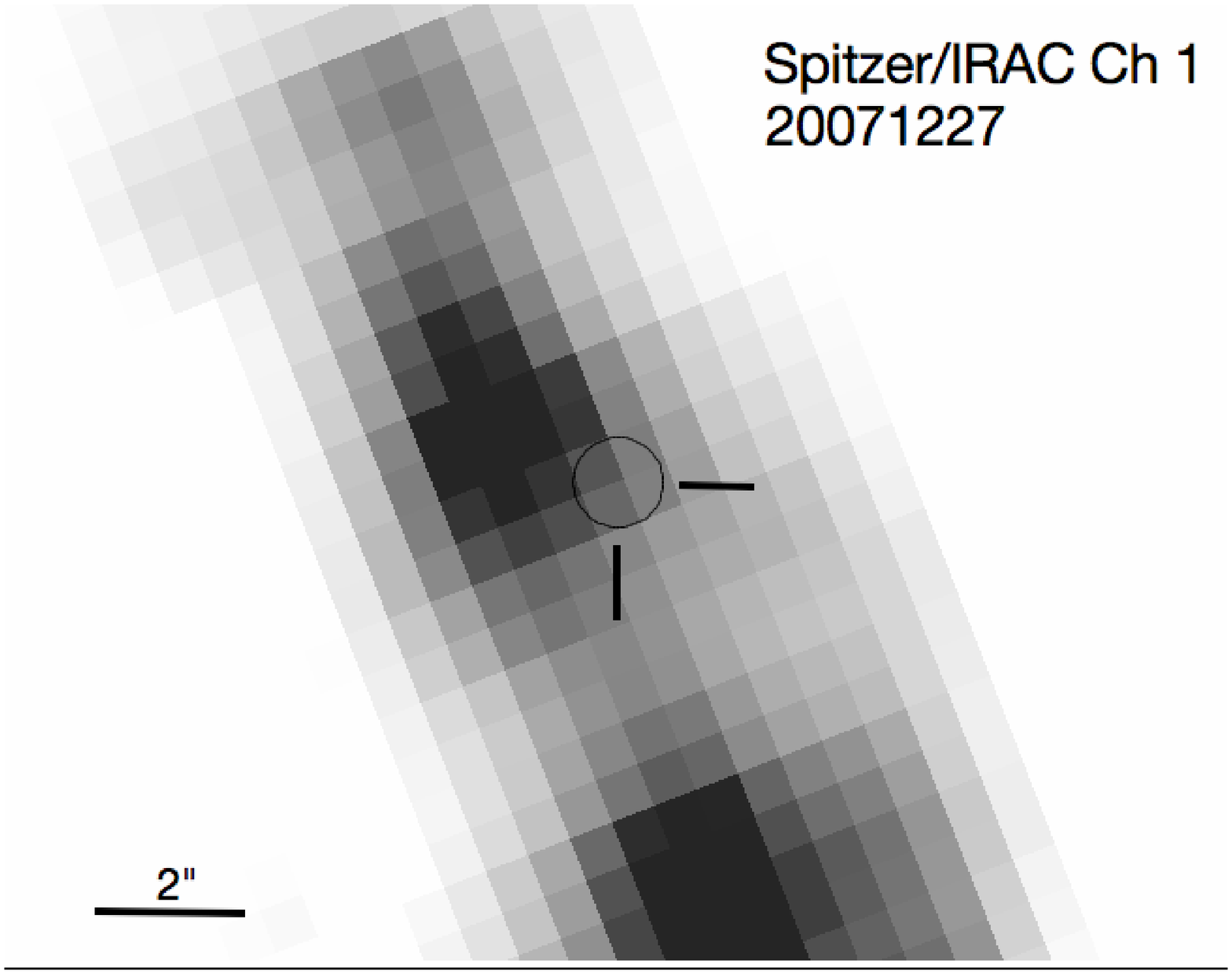}}
\subfigure{\label{f3b} \plotone{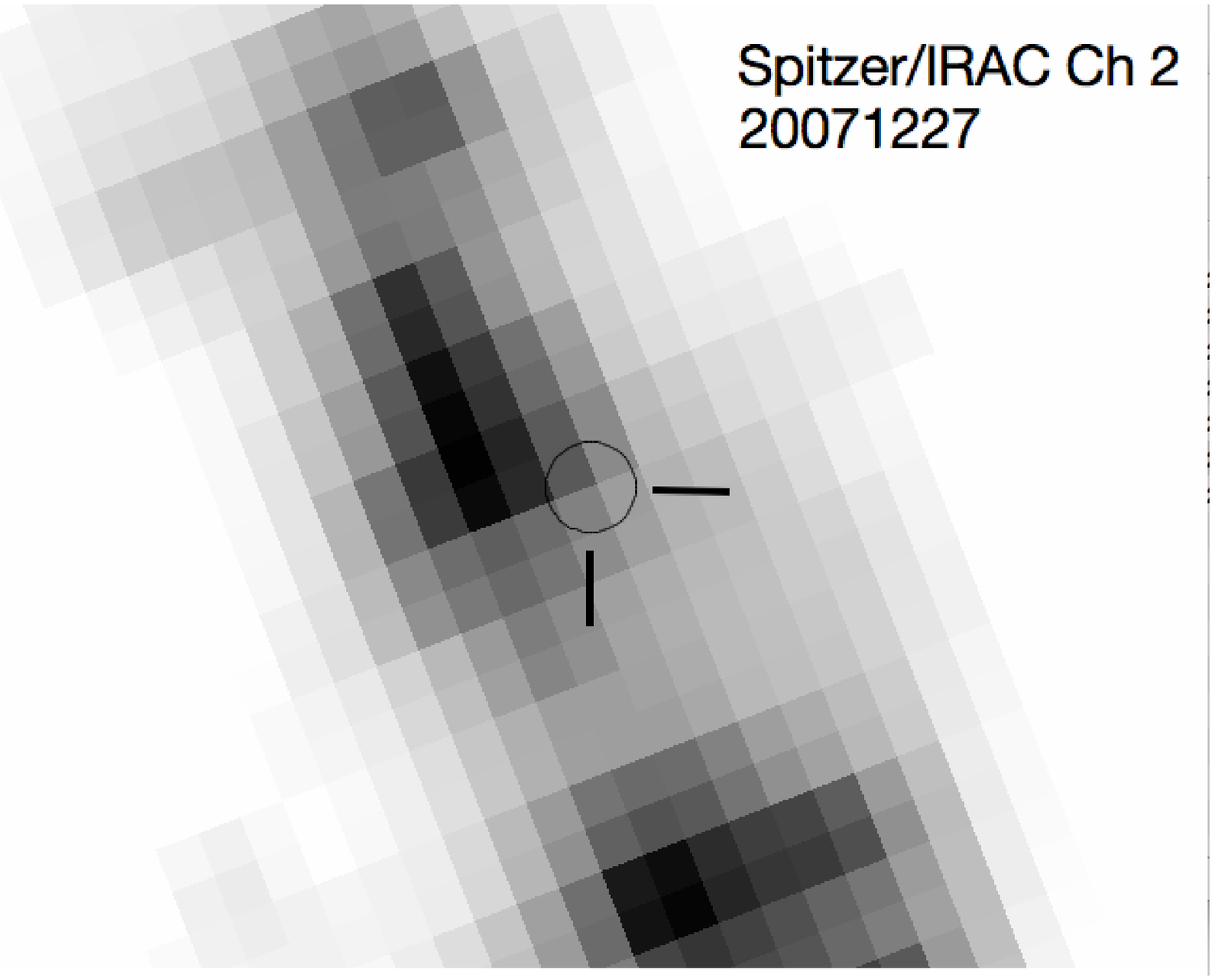}}\\
\subfigure{\label{f3c} \plotone{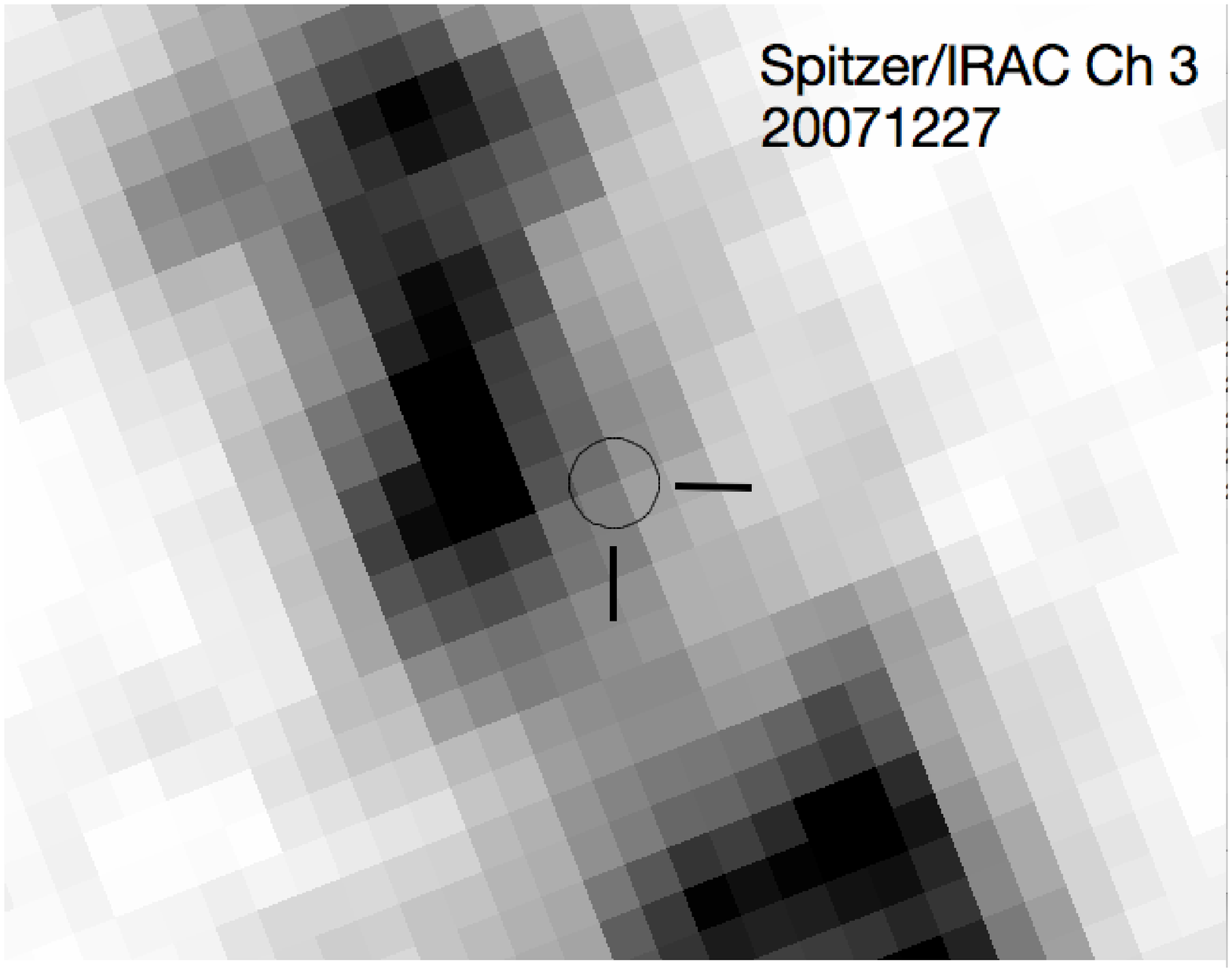}}
\subfigure{\label{f3d} \plotone{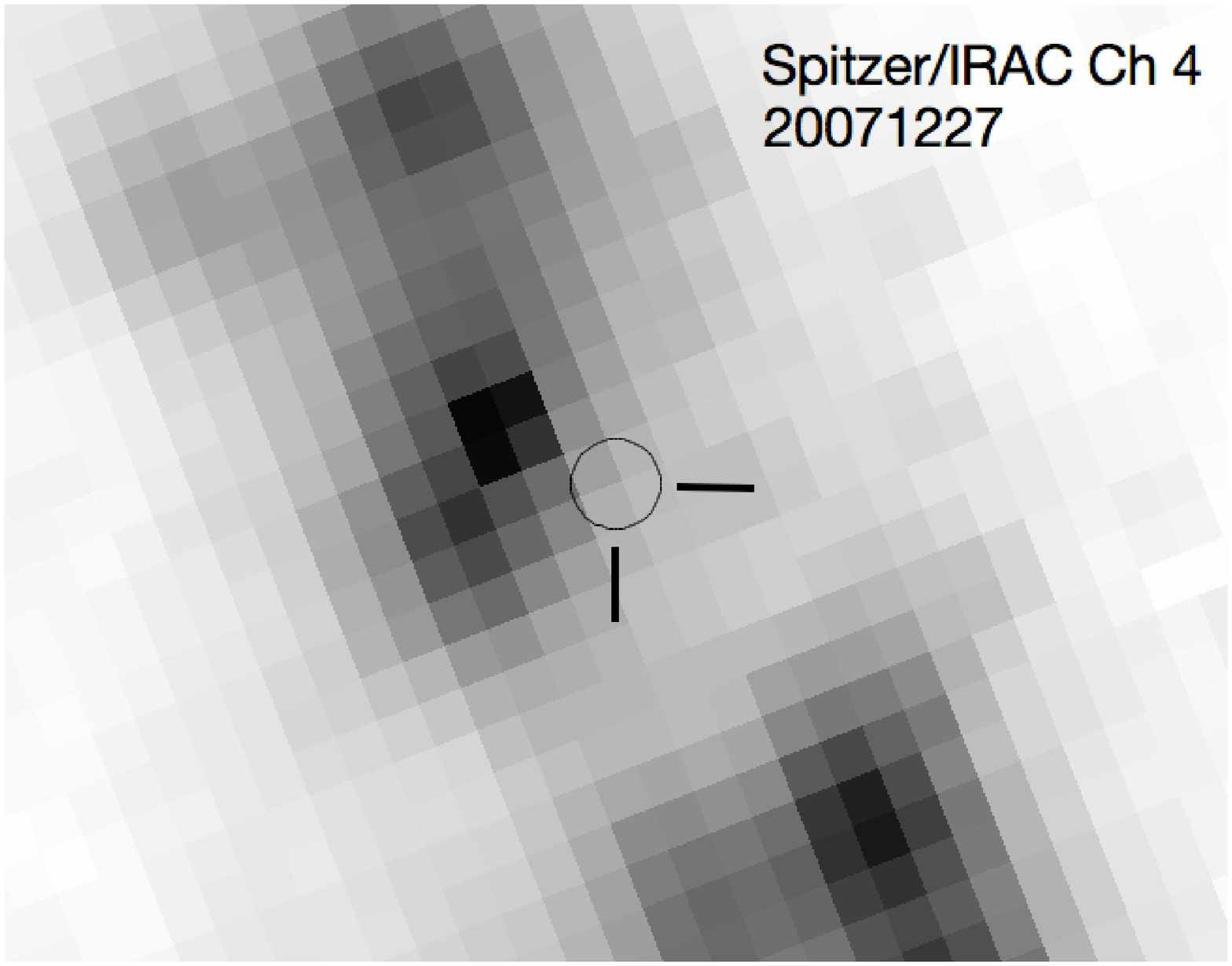}}
\caption{{\it Spitzer} Pre-Explosion data obtained on UT Dec. 27, 2007.  There is no identifiable point source at the SN position.  The lack of a detection provides useful constraints on the amount of dust surrounding the progenitor at the time of the explosion.
}
\label{fig3}
\end{center}
\end{figure*}

\subsection{The SN Position and\\ New Constraints on the Progenitor Flux}
\label{sec:astrometry}

The F336W image (Figure \ref{fig1}c) shows that SN 2010jl is offset from an extended emission region that contributes at least partially, if not entirely, to the blue object identified as the progenitor in the WFPC2 data (see Figure \ref{fig1}e; also \citealt{smith11jl}).  Figure \ref{fig2} shows a zoomed-in version.  DOLPHOT PSF-fitting routines yield the precise SN position in the 2015 WFC3/F336W image.  The SN position is centered on pixel [2152.06, 287.20]$\pm$[0.1, 0.1], while the previously identified candidate progenitor is centered on pixel [2154.52, 286.76]$\pm$[0.1, 0.1].

This offset rules out the possibility that the blue source is the star that exploded (scenarios 2 and 3 in Section \ref{sec:intro}).  The possibility that the progenitor is part of a massive star cluster (i.e., scenario 1) must still be considered.  The SN position is offset from the candidate progenitor \citep{smith11jl} in pixel space by [2.46, 0.44]$\pm$[0.14, 0.14] pixels, or $\sim2.50\pm0.2$ pixels in quadrature.  With a WFC3 scale of $0{\farcs}0396$ pixel$^{-1}$ ($\sim9.6$ pc pixel$^{-1}$ at 50 Mpc), this translates to $0{\farcs}099\pm0{\farcs}008$~($24\pm2$ pc) separation.  A typical OB association is several tens of pc across; for example, the O stars in the Carina nebula are spread across more than 40 pc \citep{smith10carina}.  This result suggests that the progenitor of SN 2010jl, even though it is not detected directly, is still likely associated with the very blue cluster.  

\citet{smith11jl} find that this blue candidate source, if not dominated by the progenitor star itself, is consistent with a young star cluster with an age of 5-6 Myr (or younger if there is host-galaxy extinction).  A single-star member of such a young star cluster reaching core collapse would be among the most massive stars in that cluster, corresponding to an initial mass of $>30$ \msolar.    

\begin{figure*}[t]
\begin{center}
\includegraphics{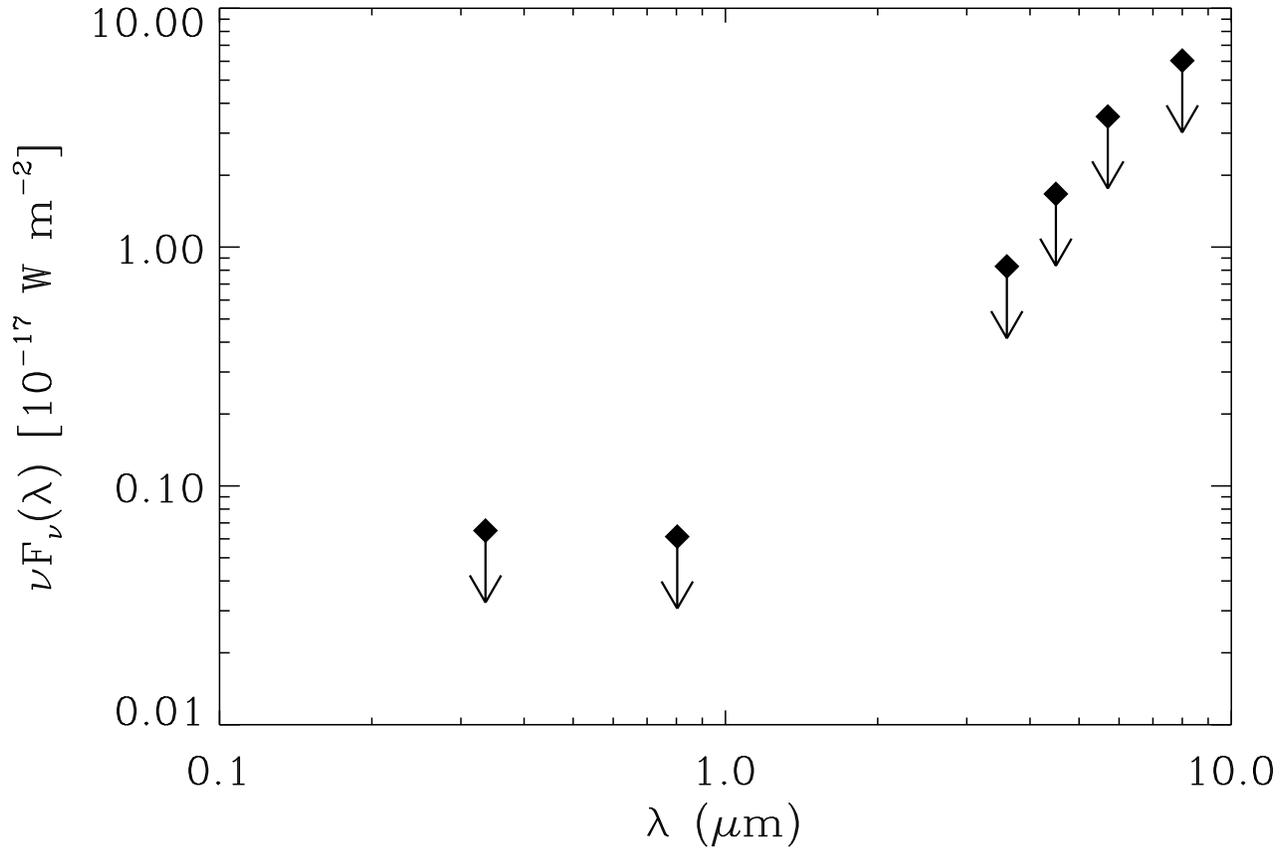}
\caption{\label{upplim} Measured ultraviolet/optical/infrared limits (black diamonds) on the fluxes from the progenitor star of SN 2010jl.  Optical upper limits are derived the pre-explosion {\it HST}/WFPC2 images, while the IR upper limits are derived from pre-explosion {\it Spitzer}/IRAC images.  These upper limits account for fluxes from both the progenitor star and any hot circumstellar dust. \label{fig4}
}
\end{center}
\end{figure*}

We determine the detection limit for any potential progenitor at the position of the SN in the pre-explosion WFPC2 images by inserting artificial stars using Dolphot v2.0 at the SN position.  To translate the SN position onto a pixel position in the 2001 WFPC2/F300W pre-explosion image, we execute the {\tt IRAF} {\tt GEOMAP} and {\tt GEOXYTRAN} commands using a list of centroids from $\sim$15 point sources identified in both images.  This analysis and alignment is completed entirely in pixel space since only relative astrometry is required.  Table ~\ref{tab1} and Figure \ref{upplim} include these limits.

\subsection{Spitzer Pre-Explosion Progenitor Constraints}
{\it Spitzer}~obtained a single epoch of observations of the host galaxy, UGC 5189A, in 2007.  We obtained fully coadded Post Basic Calibrated Data ({\tt pbcd}) from the {\it Spitzer} Heritage Archive (SHA)\footnote{http://sha.ipac.caltech.edu/applications/Spitzer/SHA/ can be used to access SHA.}, shown in Figure \ref{fig3}. We examine these images for a progenitor bright in the mid-infrared (IR), similar to the dust-enshrouded progenitor of SN 2008S \citep{prieto08}, but find no obvious point source at the SN position measured in post-explosion {\it Spitzer} data.

We determine the detection limit for any potential progenitor at the position of the SN in the pre-explosion images by inserting artificial stars using {\it Spitzer}'s MOPEX tool.  To increase our detection sensitivity for faint sources, we follow the {\it Spitzer} Help Desk's recommendation to inject artificial sources onto individual Basic Calibrated Data files (BCDs) using the APEX QA Multiframe module and then re-mosaic the data using the Overlap and Mosaic modules within MOPEX.

We utilize ``SExtractor” for our {\it Spitzer} source detection.  The narrow ``mexhat'' filter optimizes our detections since it assumes that the sources are very compact and detects them relative to the very local background.  We vary the size of the mexhat filter and set the following relevant parameters: DETECT\_MINAREA=5, DETECT\_THRESH=1.5.  We define our detection threshold as the flux of the artificial input source that we can no longer recover using the method described above.  The resulting detection limits in the four IRAC bands are as follows: 2.3$\times10^{-19}$ (Ch1), 3.5$\times10^{-19}$ (Ch2), 6.06$\times10^{-19}$ (Ch3), and 7.5$\times10^{-19}$ (Ch4)~erg s$^{-1}$~cm$^{-2}$~\AA$^{-1}$.  Figure \ref{fig4} plots these limits.

\section{Conclusion}
\label{sec:con}
Recent {\it HST}/WFC3 imaging of the SN 2010jl field obtained in 2015 shows that the SN has faded sufficiently to allow for new constraints on the progenitor.  The SN position is demonstrably offset from an underlying and extended source of emission that contributes at least partially, if not entirely, to the blue object identified as the progenitor in the WFPC2 data.  This point alone rules out the possibility that the blue source in the pre-explosion images is a single star that exploded.  

We also present previously unpublished pre-explosion {\it Spitzer}/IRAC data.  No point source is detected at the SN position.  The pre-explosion {\it HST} upper limits constrain the minimum amount of extinction required to hide a massive progenitor, while the pre-explosion {\it Spitzer}~upper limits constrain the maximum amount of flux emitted by pre-existing dust and, therefore, the {\it maximum} warm-dust mass.  (A larger reservoir of dust may exist at cooler temperatures not probed by {\it Spitzer}.)  Together, these constraints present a phase space of viable dust characteristics that could potentially extinguish a given progenitor.  We plan a future paper to analyze various dust and progenitor models within these constraints.\\
\newline

This work is based on observations made with the NASA/ESA {\it Hubble Space Telescope}, obtained at the Space Telescope Science Institute, which is operated by the Association of Universities for Research in Astronomy, Inc., under NASA contract NAS 5-26555. Support was provided by NASA through grants GO-13341 and GO-14149 from the Space Telescope Science Institute, which is operated by the Association of Universities for Research in Astronomy, Inc., under NASA contract NAS 5-26555. A.V.F.'s group is also grateful for generous financial assistance from the Christopher R. Redlich Fund, the TABASGO Foundation, and NSF grant AST-1211916. Part of the research was carried out at the Jet Propulsion Laboratory, California Institute of Technology, under a contract with NASA. E.D. acknowledges NASA's ADP13-0094 grant for support on this project.  R.J.F.\ gratefully acknowledges support from NSF grant AST--1518052 and the Alfred P.\ Sloan Foundation.  The authors would like to thank Christa Gall, Rubab Khan, Jon Mauerhan, and Arka Sarangi for their helpful discussions.  We would especially like to thank the {\it Spitzer} Help Desk at IPAC for their useful support with the MOPEX tool. 
\bibliographystyle{apj2}
\bibliography{references}

\end{document}